\documentclass[12pt]{article}   
\def\C{{\rm\kern.24em \vrule width.02em 
   height1.6ex depth-.05ex \kern-.26emC}}
\def\CC{{\rm\kern.1em \vrule width.02em 
   height1.07ex depth-.05ex \kern-.16emC}}
\def\R{{\rm I\kern-.20em R}}
\def\N{{\rm I\kern-.26em N}}
\def\Z{{\rm Z}\!\!\!{\rm Z}}

\newcommand{\be}{\begin{equation}}
\newcommand{\ee}{\end{equation}}
\newcommand{\bee}{\begin{eqnarray}}
\newcommand{\eee}{\end{eqnarray}}

\newcommand{\ie}{i.e.}
\newcommand{\del}{\partial}

\newcommand{\xx}{x_\vartheta}
\newcommand{\xxx}{y_\vartheta}

\newcommand{\ag}{\ast_\gamma}
\newcommand{\ce}{\check{\varepsilon}}

\newcommand{\qslc}{U_q\rm{sl}(2,\CC)}

\newcommand{\cX}{\hat{\chi}}
\newcommand{\RRr}{\Re}
\newcommand{\RRn}{\Re_N}
\newcommand{\Ri}{{\cal S}_1}
\newcommand{\Rin}{{\cal S}^N}

\newcommand{\RD}{\R_{\rm D}}

\newcommand{\ha}{a}
\newcommand{\ad}{a^\dagger}
\newcommand{\ath}{a_\theta}
\newcommand{\had}{\hat{a}^\dagger_q}
\newcommand{\ata}{a^\dagger_\tau}

\newcommand{\zx}{\chi}
\newcommand{\cA}{{\cal A}}
\newcommand{\cM}{{\cal M}}
\newcommand{\cL}{{\cal L}_H}
\newcommand{\cT}{{\cal T}_2}
\newcommand{\cTT}{{\cal T}_1}
\newcommand{\cLL}{{\cal L}_H^q} 
 
\newcommand{\cLLs}{ {\cal L}_H^q }
\newcommand{\cLLq}{\widehat{\cal L}_q}
\newcommand{\cB}{\varpi}

\newcommand{\cI}{{\cal I}}

\newcommand{\vR}{{}^\vee\R}

\newcommand{\cAA}{\widehat{{\cal A}}  }
\newcommand{\D}{\widehat{{\cal O}}}
\newcommand{\Dc}{\widehat{O}}
\newcommand{\Dg}{{\cal D}^\gamma}
\newcommand{\Dq}{{\cal D}^q}
\newcommand{\dd}{ {\rm d}_q  }
\newcommand{\cR}{{\cal R}}
\newcommand{\cD}{{\cal D}}

\newcommand{\cF}{\Re}
\newcommand{\cS}{{\cal S}}
\newcommand{\zz}{\chi}

\newcommand{\qRN}{\R^N_q}
\newcommand{\llmapsto}{
   \,\, \rule[.5pt]{.5pt}{1.63mm}\hskip-.11cm \longmapsto \, \,
    }
\newcommand{\mmapsto}{
  \,\, \rule[.5pt]{.5pt}{1.63mm}\hskip-.13cm \mapsto 
    }

\newcommand{\cx}{\check{x}}
\newcommand{\cp}{\check{+}}
\newcommand{\cep}{\check{\epsilon}}

\makeatletter
\begin{document}
\baselineskip.7cm


\vspace{-.2cm}
\vspace{1.0cm}
\begin{center}
\LARGE{ 
      Quantizations of $\R$ 
}\\[2.3em]
\large{  Takashi Suzuki
}\\[.3cm]     
\normalsize    Hiroshima Institute of Technology            \\
\normalsize   Hiroshima 731-5143,  JAPAN\\
 \normalsize  e-mail : stakashi@cc.it-hiroshima.ac.jp

\end{center}           
\makeatother

\vspace{.15cm}

\begin{abstract}
Quantum real numbers are proposed 
 by performing a quantum deformation of the standard 
 real numbers $\R$. 
We start with the $q$-deformed Heisenberg algebra $\cLLq$ 
which is obtained by the Moyal 
$\ast$-deformation of the Heisenberg algebra 
generated by $a$ and $\ad$.   
By representing  $\cLLq$ as 
the algebras of $q$-differentiable functions,   
we  derive quantum real lines 
from the base spaces of these functional algebras.  
We find that these quantum lines are discrete spaces. 
In particular,  for the case with 
$q =  e^{2\pi i \frac{1}{N} }  $, 
the quantum real line  is composed of 
fuzzy, \ie, fluctuating  points  
and nontrivial infinitesimal structure appears 
around every standard real number.   
\end{abstract}

\section{Introduction}

The long-standing problem, quantization of 
gravity theory, 
undoubtedly requires some fundamental modifications 
of the standard theory of geometry.
By taking quantum effects into account, 
the concepts of, e.g., point, line, and so on 
should be reconsidered:  
(\romannumeral1) 
What is a point? 
Is it just a localized object of null size?
(\romannumeral2) What is a line? 
Is it just a one-dimensional continuous object?  
In other words, we should ask what quantum real number $\R$ is. 
The aim of this paper is to challenge 
these essential problems.

As for the first problem, it can be expected that 
a point suffers quantum fluctuation and becomes something fuzzy, \ie, nonlocal object with some extra structure.
Let us call such a  quantum object a {\it quantum point}.
As a set of quantum points, 
a {\it quantum real line} can be defined and, 
therefore, quantum real line must be fuzzy.
{
Although we are much interested in geometries  
of higher dimensional quantum objects, 
this paper focuses on the quantum real line 
as the first step towards the quantum geometry.

A well-established example of quantum geometry 
is provided by the noncommutative geometry 
developed by Connes \cite{Co}.
The noncommutative geometry has appeared 
in the recent developments in string theory.
Actually, noncommutative coordinates are used to 
describe D-brane.
Another example is the quantum geometry based on 
quantum groups (see, for example, Ref.\cite{Ma}).
Let us observe briefly the above two approaches 
of quantization of geometry.  
Let $M$ be the classical manifold or space  
which is to be quantized. 
The key process in both approaches 
is to define a commutative algebra $A(M)$ on $M$. 
One then deform $A(M)$ to a noncommutative algebra 
$\cA(\cM)$ by some deformation procedure.   
In the Connes' noncommutative geometry, 
a functional algebra on $M$ is taken as 
$A(M)$, the algebra to be deformed.
As an extension of the Gelfand-Naimark theorem. 
which guarantees that geometrical data of $M$ are  
extracted from $A(M)$,  
noncommutative geometry of the quantum space $\cM$ 
is derived from the noncommutative algebra $\cA(\cM)$.
On the other hand, in the quantum group approach, 
$A(M)$ is provided by the coordinate ring $A(G)$ of the group 
$G$ which acts on the space $M$. 
The quantum geometry of the quantum space $\cM$ 
is derived from the quantum group $A_q(G)$.

Now,  we are interested in the step 
from a commutative algebra $A$ 
to a noncommutative algebra $\cA$.
In the quantum geometry via the quantum group, 
the quantum algebra $A_q(G)$ is obtained from $A(G)$ 
according to the general ideas of quantum group.  
In the Connes' approach, however, 
one can reach $\cA(\cM)$ by  
some procedures. 
A possible way is the Moyal deformation quantization,   
\ie, as a multiplication structure of $\cA(\cM)$,    
a noncommutative Moyal $\ast$-product is used  
instead of the standard  commutative  product \lq\lq$\cdot$" in $A(M)$.   
Let us explain in more detail.  
Let $\gamma$ be the deformation parameter 
which brings $A(M)$ to $\cA(\cM)$. 
The algebra $\cA(\cM)$ possesses the multiplication 
$\ag$ which is associative but noncommutative, 
\ie, for functions $f,g,h \in \cA(\cM)$, 
$(f\ag  g) \ag h = f \ag (g \ag h)$ and 
$f \ag g \neq g\ag f$. 
Note here that elements of $\cA(\cM)$ 
are  the same as those of $A(M)$, and   
the only difference between $A$ and 
$\cA$ is in the multiplication structure.
Since geometrical data of $\cM$  are deduced from $\cA(\cM)$,   
$\cM$ is regarded as a noncommutative manifold.  
In particular, therefore, the coordinates in $\cM$ 
which are elements of $\cA(\cM)$ 
do not commute because of $\ag$. 
By taking the limit $\gamma \rightarrow 0$, 
$\cA(\cM)$ reduces to $A(M)$ and, therefore,   
 $ \cM \stackrel{\gamma\rightarrow0}{\longrightarrow} M$.


In this paper, we will propose quantum real numbers 
and show the geometries of  quantum real lines     
guided partly by the above strategy 
and partly by Ref..\cite{TS1}. 
The algebra with which we will start is 
the Heisenberg algebra $\cL$ 
generated by the operators $a$ and $a^\dagger$,
since $\cL$ is equivalent to the algebra of 
differentiable functions on the real numbers $\R$. 
In order to obtain quantum real numbers, 
we will deform $\cL$ in Section 2 
and reach the $q$-deformed Heisenberg algebra 
$\cLLq$. 
The quantization from $\cL$ to $\cLLq$ is performed according to the procedure developed in \cite{TS1}. 
Namely,  $\cLLq$ is introduced as 
$\cLLq:= \cL \times \widehat{\cA}(\cTT)$.     
Here, $\widehat{\cA}(\cTT)$ is the  internal sector 
and, explicitly, an operator algebra 
on the one-dimensional torus $\cTT$. 
The operator algebra $\widehat{\cA}(\cTT)$ is derived 
from the noncommutative functional algebra $\cA(\cT;\ag)$ 
on the two-dimensional torus $\cT$. 
In Section 3, we will represent $\cLLq$ 
as the algebra of $q$-differentiable functions. 
We will, in Section 3.1, look at the case 
where $q$ is not a root of unity 
and construct the algebra $\Dq(\Re)$ of $q$-differentiable 
functions on the base space $\Re$.   
Section 3.2 looks at the case 
where $q$ is the $N$-th root of unity and 
the algebra $\Dq_N(\Re_N)$ is built.   
The section 4 is the main part of this paper, 
where the structures of the base spaces 
$\Re$ and $\Re_N$ are investigated 
through the algebras $\Dq(\Re)$ and $\Dq_N(\Re_M)$, 
respectively.  
We will finary propose, in Section 4.1, 
the quantum real line $\R_{\rm D}$ 
via the base space $\Re$.  
and, in Section 4.2,   
the quantum real lines $\qRN$ and $\vR$ via $\Re_N$. 

We will further discuss 
geometrical  features of these lines 
and show the  followings;    
the line $\R_{\rm D}$ is a discrete space 
which is composed of an infinite number of 
{\it standard} points. 
Namely, the minimal length appears in $\R_{\rm D}$. 
The lines $\qRN$ and $\vR$ are also discrete spaces. 
However, each point in these lines is not a standard point 
but a {\it fuzzy} point. 
Indeed,  we will give a definition of 
a \lq\lq wave function of a point"
 and finds that wave functions of fuzzy points  are 
not localized but spread. 
It should be remarked that 
the standard real line $\R$ appears as a subset, 
\ie, $\R \subset \qRN, \, \vR$.  
 The spaces $\qRN\setminus\R$ and $\vR\setminus\R$ 
are called {\it infinitesimal structures} 
in the sense that 
the quotient  spaces consist of points living  between $x$ and 
$x+\epsilon$ for ${}^\forall x\in\R$ and $\epsilon\in \R$ 
being an infinitesimal.

Concluding remarks, 
especially on the features in the limit $q \rightarrow 1$, 
and future problems are 
addressed in Section 5.

\section{Deformation of the Heisenberg algebra }

We start with a deformation of 
the Heisenberg algebra $\cL$,   
which is the operator algebra generated by the two generators 
$\ha$ and $\ad$ satisfying the defining relation
\be 
\ha \ad  - \ad \ha = 1.
\label{Heisen}    
\ee
Although some deformations of $\cL$ have 
been discussed in lots of works, 
let us follow the deformation procedure developed 
in Ref.\cite{TS1}, 
where the origin of quantum effects is given manifestly.

The first and essential step  
is the introduction of the algebra 
$\cA(\cT)$ of functions on the two-dimensional torus $\cT$. 
Let $\theta$ and $\tau$ be the variables on $\cT$. 
Then, the generators of $\cA(\cT)$ are  written as
\be
U:= e^{i\theta}, \qquad V := e^{i \tau}, 
\label{geneL}    
\ee
\ie, $\cA(\cT)\ni f=f(U,V)$. 
Upon interpreting $\cA(\cT)$ as the internal sector, 
we will take the product $\cL \times \cA(\cT)$ later  
in order to  define our deformed Heisenberg algebra.  
Before doing this,
we have to introduce a multiplication into $\cA(\cT)$.  

The most familiar and simplest one  
is the standard local product \lq\lq $\cdot"$ 
and let $\cA(\cT;\cdot)$ be the algebra 
with $\cdot\,$.       
Obviously, $\cA(\cT;\cdot)$ is 
a closed and commutative algebra, \ie,  
for ${}^\forall f(U,V),g(U,V)\in \cA(\cT;\cdot)$, 
$f \cdot g = g\cdot f \in \cA(\cT;\cdot)$. 
In this case, however, 
one {\it cannot} expect 
$\cL \times \cA(\cT;\cdot)$ to be a nontrivial deformation 
of the Heisenberg algebra.  
Indeed, one concludes with the isomorphism 
$\cL\times \cA(\cT;\cdot) \cong \cL$,      
since the internal sector $\cA(\cT;\cdot)$ 
is factorized completely from the algebra $\cL$.  

Another possible product is the, so-called, 
Moyal product $\ast_\gamma$ 
with a deformation parameter $\gamma$.   
Let $\cA(\cT;\ag)$ be the algebra 
endowed with $\ag$. 
Here, the Moyal product $\ast_\gamma$ 
is given explicitly as  \cite{TS1} 
\bee
\ag & = & \sum_{n=0}^\infty  \frac{
                   \left( \frac{-i\gamma}{2}\right)^n} {n!}
\sum_{k=0}^n \left(\begin{array}{c} n \\ 
   k \end{array}\right) (-)^k 
    \overleftarrow{\partial_\theta^{\,n-k}\partial_\tau^k}
      \cdot 
    \overrightarrow
    {\partial_\theta^{k}\partial_\tau^{\,n-k}}.     
\label{gproduct}    
\eee
Since the product  (\ref{gproduct}) is nonlocal,  
the algebra  $\cA(\cT;\ag)$ is  
no longer commutative algebra but 
a noncommutative algebra 
with the defining commutation relations as 
\bee
  & U \ag V= e^{i\gamma} V \ag U,& \label{uv}   \\[.13cm]
  & \theta \ag \tau  - \tau \ag \theta =-i\gamma. &  
\label{crxiphi}    
\eee
The commutation relations in (\ref{uv}, \ref{crxiphi}) 
indicate that the base space $\cT$ of 
$\cA(\cT;\ast_\gamma)$ is 
the so-called noncommutative torus. 
into the Heisenberg algebra. 
Let us define 
\be
\cLL= \cL \times \cA(\cT;\ast_\gamma).
\ee
One can now expect that $\cLL$ 
 provides nontrivial deformations  of $\cL$.  
Let us explain in more detail. 
We define the algebra $\cLL$ explicitly by giving the generators 
as 
\be
  \ath := \ha U,\qquad   
  \ata := \ad V. 
\label{defgene}   
\ee
Then, the algebra $\cLL$ 
is an operator algebra  
spanned by the functions of operators $\ath, \, \ata$, 
where the operator nature originates only in the sector $\cL$.  
Tthe multiplication is introduced into $\cLL$ as follows; 
the standard operator product is assumed in the sector $\cL$, 
while the Moyal product (\ref{gproduct}) is used 
for the internal sector $\cA(\cT;\ast_\gamma)$. 
By making use of the commutation relations given in 
eqs.(\ref{Heisen}) and (\ref{uv}),  
one immediately obtains the deformed commutation relation 
between the generators $\ath,\,\ata$ as    
\be
   \ath \ag  \ata- q\,\, \ata \ag \ath = 
                \, U\ag V, 
 \label{qcr} 
\ee
where $q$ is the deformation parameter given by
\be
   q = e^{i\gamma}. 
\label{qvalue}    
\ee
Note here that, in the limit $\gamma \rightarrow 0$, 
the deformed algebra $\cA(\cT;\ag)$ reduces to 
the commutative algebra $\cA(\cT;\cdot)$. 
Therefore, as discussed in the preceding paragraph, 
the deformed algebra $\cLLs$ reduces to 
the original Heisenberg algebra $\cL$,  \ie, 
$\cLLs \stackrel{\gamma\rightarrow 0}{\longrightarrow} \cL$.

Having obtained the operator algebra $\cLLs$, 
our interests are in a representation 
of the algebra and the geometry of the base space 
on which $\cLLs$ acts. 
According to the Gelfand-Naimark  theorem 
and the strategy of Connes \cite{Co}, 
the geometrical structures of the space 
will be extracted from $\cLLs$. 
These investigations are the tasks 
in the following sections.

However, a formulation with the Moyal product is 
not always convenient to handle. 
Let us, therefore, derive another $1$-deformed 
algebra without the Moyal product  from $\cLLs$.  
In the derivation, we are  required that 
the $q$-deformed  effects  originated from 
the relation (\ref{qcr}) should be preserved. 
Recalling  that, in $\cLL$,  the $q$-deformations come from 
the noncommutativity   
of $\cA(\cT;\ag)$,  
our problem is how to derive the noncommutativity  
without using the Moyal product. 
A solution is to use the operator formalism as follows: 
The functional algebra $\cA(\cT;\ag)$ is changed into 
the operator algebra $\cAA(\cTT)$ 
where $\cTT$ is a one-dimensional configuration space 
reduced from $\cT$ regarded as  the phase space. 
Namely, $\cAA(\cTT)$ acts on the space of functions on $\cTT$. 
We introduce the operator-valued generators 
$\hat{U},\, \hat{V} $ of $\cAA(\cTT)$ so that 
they should  satisfy the commutation relation 
\be
\hat{U} \hat{V} = q \hat{V} \hat{U}
\label{quv}   
\ee
instead of (\ref{uv}). 
Upon writing these generators as  
$\hat{U}=e^{i\hat{\theta}},\, \hat{V}=e^{i\hat{\tau}}$, 
one easily finds that  
the requirement given in (\ref{quv}) follows from 
the commutation relation 
\be
  \hat{\theta}\hat{\tau} - \hat{\tau}\hat{\theta} 
   = -i\gamma, 
\label{Intcom}    
\ee
which is compatible with the relation  
(\ref{crxiphi}). 
Then, a representation of the algebra  $\cAA(\cTT)$ 
is performed 
by imposing the polarization, 
\be  
\hat{\theta} = \theta, \qquad 
\hat{\tau}= i\gamma \frac{d}{d\theta} := 
    i\gamma \del_\theta.   
\label{phi}     
\ee 
Therefore, 
the configuration  space  $\cTT$ is one-dimensional torus 
parameterized by $\theta,\, 0 \leq \theta < 2\pi$.

Now, we have 
obtained the internal sector $\cAA(\cTT)$.  
Llet us define the $q$-deformed Heisenberg algebra  $\cLLq$ as 
\be 
  \cLLq := \cL \times \cAA(\cTT).  
\label{defcTT}    
\ee
What we should make explicit are 
the generators of $\cLLq$ 
and the defining commutation relation between them. 
Let $\hat{\ha}_q,\, \had$ be the generators 
and  define them by the following substitutions, 
\be
  a \hat{U}\,\longmapsto\,\hat{\ha}_q, \qquad 
  e^{i\theta}\,\ad \hat{V}\,\longmapsto\, \had.  
\label{changes}   
\ee
The defining relation of $\cLLq$ is finally obtained as
\be
\hat{\ha}_q \had -q\, \had \hat{\ha}_q = q^{i\del_\theta}.
\label{newcr} 
\ee

\section{Deformations of algebra of functions on $\R$}

Recall first that a possible 
representation of the Heisenberg algebra $\cL$ 
is realized by the replacements $\ha\rightarrow \del_x$ and 
$\ad \rightarrow \hat{x}$ 
where $\del_x f(x) = \frac{d}{dx} f(x)$ and 
$\hat{x}f(x)=xf(x)$ 
for differentiable  functions $f(x)$ on $\R$. 
We denote the operator algebra generated by $\hat{x}$ 
and $\del_x$ as $\Dc(\R)$. 
Let us adopt similar replacements to 
the $q$-deformed Heisenberg algebra $\cLLq$ 
and obtain the $q$-deformed operator algebra $\D(\cF)$ 
which acts on the space of  functions on the space $\cF$.  
Upon taking the relationship between $\cL$ and $\R$ 
into account, 
we expect that $\cF$ is raised to  a $q$-deformation of $\R$. 
We should give a comment here on the base space $\cF$. 
Although, according to our construction of $\cLLq$, 
it is natural to regard $\cF$ as the tensor product of 
two spaces, \ie,  
the space for $\cL$ and just $\cTT$, 
we will consider $\cF$ to be the one-dimensional 
space parameterized by $\vartheta$ as
\be
  \cF=\{ \vartheta \,\vert  
     -\infty \leq \vartheta \leq \infty \}, 
\ee
and find such a representation.

Let $\cX$ and $D_q$ be the 
generators of $\D(\cF)$. 
Namely, upon the replacements 
\be
{}\hskip-.5cm 
\hat{\ha}_q \,\longrightarrow \, D_q, \qquad 
\had\, \longrightarrow \, \cX,
\label{replace}    
\ee
$\cX$ and $D_q$ are identified, respectively,  with  
the $q$-deformed coordinate operator and 
the $q$-deformed differential operator. 
Let $\Dq (\cF)$ be the algebra of functions
on which the action of $\D(\cF)$ is defined. 
Geometrical structure of $\cF$ 
is deduced from $\Dq(\cF)$.  
In the following subsections, we are going to 
investigate $\Dq(\cF)$ for two cases;  
Section 3.1 looks at the case with a generic $q$ 
and the case with $q$ at a root of unity 
is studied in Section 3.2.

\subsection{The case  where $q$ is generic}

Let us suppose first that the deformation parameter 
$q$ is not a root of unity.
Upon the substitutions (\ref{replace}), 
the defining relation (\ref{newcr}) reads 
the commutation relation between $D_q$ and $\cX$ as
\be
D_q \cX - q\, \cX D_q = q^{i\del_\vartheta}.
\label{com}     
\ee
The explicit expressions of these operators are given  by
\be
\cX =  \frac{1+ q^{i\del_\vartheta}}{1+q^{-1}} \, 
                           e^{i\vartheta}, \qquad \quad
D_q = \frac{q^{-i\del_\vartheta}-1}
                       {e^{i\vartheta}(q-1)}. 
\label{defaa}   
\ee
Let us use the variable $\xx$ instead of $\vartheta$  
and redefine the base space $\cF$ as
\be
  \RRr=\{ \xx \,\vert \,
     \xx := e^{i\vartheta}, \,\,  
     -\infty\leq \vartheta \leq \infty  \}. 
\label{qcoordinate}     
\ee
One then constructs the functional algebra  $\Dq(\RRr)$ as 
\be
   \Dq(\RRr) =\{ f(\xx)\,\, \vert \,\, \xx\in \RRr,
     \},  
\ee 
where the multiplication in $\Dq(\RRr)$ 
is supposed to be the standard 
pointwise product \lq\lq $\,\cdot$".
The complete basis of $\Dq(\RRr)$ are given by 
\be
     \cB := \{\xx^n\,\,\vert \,\, n\in \Z\} 
\label{basis}    
\ee
and, therefore, ${}^\forall f(\xx) \in \Dq(\RRr)$ 
is expanded as
\be 
f(\xx) = \sum_{n\in \Z} f_n\, \xx^n .
\ee
The actions of the operators $\cX$ and $D_q$ 
on $\Dq(\RRr)$  are calculated explicitly by (\ref{defaa}) as     
\be
\cX\, f(\xx) = \xx \, 
     \frac{f(\xx) + q^{-1}f(\xx q^{-1})}{1+q^{-1}}, \qquad
D_q f(\xx) =
           \frac{f(\xx q) -f(\xx)}{\xx(q-1)}.
\label{defxiD}    
\ee
Note that, by taking the limit $q\rightarrow 1$ 
together with $\xx \rightarrow x$, 
$\cX$ becomes the standard coordinate operator  
$\hat{x}$ 
and $D_q$ reduces to the standard differential operator 
$\del_x$, \ie,  
$\D(\RRr) \ni \cX, \, D_q 
    \stackrel{q\rightarrow}{\longrightarrow} 
    \hat{x}, \, \del_x \in \widehat{O}(\R)$. 
Furthermore, one shows that $D_q$ 
satisfies the following properties,   
\bee
\mbox{}\hskip-1cm
&& D_q \, {\bf 1} =0, 
   \quad  \mbox{for the identity}\,\,{\bf 1}\in\Dq(\RRr),  
             \label{constant}        \\[.15cm] 
 && 
    D_q \left( f\cdot g \right) = 
     \left( D_q f(\xx) \right) g(\xx) 
     + f(q\xx) \left( D_q g(\xx) \right). 
  \label{Leibniz}    
\eee 
The first equation (\ref{constant}) indicates that 
$D_q$ vanishes constants in $\Dq(\RRr)$, 
and (\ref{Leibniz}) shows the deformed Leibniz rule.  

It is interesting to observe  finally that 
the $q$-deformed exterior derivative $\dd$ can be 
defined  through 
\be
\dd f  = \dd \xx D_qf(\xx).  
\ee   
One immediately finds that\be
\dd(f \cdot g) = (\dd f) \cdot g + f \cdot (\dd g)
\ee   
holds if and only if another noncommutativity\be
\dd \xx\,f(q\xx) = f(\xx)\,\dd \xx
\label{NC}   
\ee 
is satisfied. 
Thus we have found the \lq\lq noncommutative"  structures 
shown in eqs.(\ref{Leibniz}, \ref{NC}). 
This is the reason why we will refer to $\Dq(\Re)$ 
as \lq\lq noncommutative algebra", 
although any elements in $\Dq(\Re)$ 
are mutually commutable. 
In other words, the noncommutative natures 
originate just from the $q$-differential operator.

\subsection{The case where $q$ is a root of unity}

Let us  turn to the case where the deformation parameter 
$q$ is a root of unity, 
\ie, for a positive integer $N$,  
\be
       q = e^{i\,\frac{2\pi}{N}}.
\label{rootofunity}     
\ee
We will use the variable $\xxx$ in this case 
and rewrite the base space $\cF$ as 
\be
\RRn =\{\, \xxx\, \vert \, \xxx := e^{i\vartheta} , \,\,
      -\infty \leq \vartheta \leq \infty \}.   
\label{qqline}     
\ee
Let $\Dq_N(\RRn)$ be the algebra of 
functions 
on which the $q$-coordinate operator $\cX$and the 
$q$-differential operator $D_q$ 
with respect to $\xxx$ are defined. 
The basis of $\Dq_N(\RRn)$ are then 
\be
\cB_N:=\{\,\xxx^n\,\vert\, n\in \Z\,\},  
\label{baseN}   
\ee
\ie, any function $f(\xxx)\in \Dq_M(\RRn)$ 
is expanded as 
\be
  f(\xxx) = \sum_{n\in \Z} \alpha_n\,\xxx^n. 
\label{defImqN}     
\ee
Before going to the investigations of $\Dq_N(\RRn)$, 
recall that representations of a quantum group $G_q$ 
with $q$ at a root of unity are 
drastically different from those of the quantum group  $G_q$ 
with generic $q$  
and of the classical group $G$.   
We will see below that the structure of $\Dq_N(\RRn)$ is also 
different from that of 
$\Dq(\RRr)$ completely.

In order to observe the structure  of $\Dq_N(\RRn)$, 
one should start with the actions of 
the operators $\cX$ and $D_q$ defined in 
(\ref{defaa}).  
In particular, the actions of $D_q$  
on the basis $\cB_N$ are to be remarked. 
Actually, upon using $[N]=0$, one finds
\bee
D_q \xxx^{kN+r} &=& [r]\,\xxx^{kN+r-1}, \quad \mbox{for} 
      \quad r \neq 0, \nonumber \\[.15cm]
 D_q \xxx^{kN} &=& 0.
\label{operationD}  
\eee
Thus, 
$\xxx^N$ behaves as a constant with respect to $D_q$.
Further, one finds easily that 
the operator $D_q^N$ on $\Dq_N(\RRn$ 
is null, \ie, 
$D_q^Nf(\xxx)=0, \, {}^\forall f(\xxx)\in \Dq_N$.   
However, upon defining another operator $\del$ by 
$\del := D_q^N/[N]!$, 
$\del$ acts on $\cB_N$ as  
\bee 
\del\, \xxx^{kN+r} = k\,\xxx^{(k-1)N+r}, \qquad 
      && \mbox{for} \quad  k \in \Z,  \nonumber \\[.15cm]
\del\, \xxx^r =0, \qquad \qquad \qquad 
      && \mbox{for} \quad  r\leq N-1,  
\label{Naction}    
\eee
where use has been made of the relation $[kN]/[N]=k$.  
One should notice that, with respect to $\del$,  
$\xxx^r$ for $1\leq r\leq N-1$ are regarded 
as constants in $\Dq_N(\RRn)$. 
The facts derived from  
(\ref{operationD}) and (\ref{Naction}) are:
\begin{enumerate}  
\item  
These operators $D_q$ and $\del$ commute with  
each other. 
\item The variables $\xxx^{\pm r}\, (r=1,\cdots, N-1)$ 
and $\xxx^{kN},\, k\in \Z$ are 
independent of each other. 
\end{enumerate}  
Therefore, one should require that 
two independent variables $\xxx$ and $\, \xxx^N$ are needed 
for the perfect 
definition of  the algebra $\Dq_N(\RRn)$ 
and further that  
two independent ($q$-deformed) differential operators 
$D_q$ and $\del$ should be introduced.

Guided by the above observations, 
let us investigate the structure of $\Dq_N(\RRn)$ in detail.  
To this end, we introduce the map 
$\pi:\Dq_N (\RRn)  \rightarrow 
\cD(\cR) \otimes \Dq_I(\Ri)$
by
\be
\pi(\xxx^{kN+r}) = x^k\xi^r.  
\ee
Namely, for a function $f(\xxx) \in \Dq_N(\RRn)$, 
\bee
   && \pi(f(\xxx)) = \lambda(x) \otimes  \psi(\xi), \\[.15cm]
   &&{}\quad \lambda \in \cD, \quad x \in \cR \qquad 
        \psi \in \Dq_I, \quad \xi \in  \Ri,  
                      \nonumber 
\eee
where $k\in\Z$ and $r$ is in the region ${\cal I}$ as  
\be
   r \in {\cal I} := \left\{ 
  \begin{array}{ll} 
      \{ -p , -p+1, \cdots , p-1, p \},
     & \mbox{for} \,\,\, N=2p+1 , \\[.2cm] 
      \{-p+1, \cdots, p-1, p \},    & \mbox{for} \,\,\, N=2p
      \end{array}  \right. .
\label{region}     
\ee
The basis $\cB_N$ given in (\ref{baseN}) is 
also factorized as 
$\cB_N \rightarrow \cB_{out} \otimes \cB_{int}$ where 
$\cB_{out} := \{
   x^k\,\vert\, k\in \Z\}$ and  
$ \cB_{int} := \{  \xi^r\,\vert\, r \in {\cal I} \}$.   
Namely,  functions $\lambda(x) \in \cD(\cR)$ 
and $\psi(\xi)\in \Dq_I(\Ri)$ are expanded as   
\be
  \lambda(x) = \sum_{k\in \Z} \lambda_k\, x^k, \qquad  
  \psi(\xi) = \sum_{r\in {\cal I}} \psi_n \xi^r. 
                 \label{expansion1} 
\ee   

Let us go back to the ($q$-deformed) differential operators 
$\del, \, D_q$ 
and discuss their factorizations induced from $\pi$. 
The equations shown in  (\ref{operationD},\ref{Naction}) 
suggest that the operators $\del$ and $D_q$ are factorized as
\be
  \del\,\longmapsto \, \del_x\otimes id, \qquad 
  D_q \,\longmapsto \, id \otimes \nabla_q, 
\ee 
\ie, $\del_x$ and $\nabla_q$ are ($q$-)differential operators 
defined, respectively, on $\cD(\cR)$ and $\Dq_I(\Ri)$ 
such as
\be
  \del_x\lambda (x) = \frac{d}{dx} \, \lambda(x), \qquad 
  \nabla_q\,\psi(\xi) = \frac{\psi(q\xi)-\psi(\xi)}{\xi(q-1)}.
\label{dif}     
\ee

In order to make our story complete, 
we have to determine the multiplication structures 
in both sectors. 
The first sector $\cD(\cR)$ can be endowed with 
the standard multiplication \lq\lq \,$\cdot\,"$ 
and, with the multiplication, 
$\cD(\cR)$ is the standard algebra 
of differentiable functions on $\cR$. 
On the contrary, 
the sector $\Dq_I(\Ri)$ cannot possess \lq\lq \,$\cdot\,"$, 
since the sector does not close 
with respect to the product.
Actually, for $r,\,r' \in \cI$, the value $s$ such as 
$\xi^r \cdot \xi^{r'}=\xi^s$ is not always in $\cI$.
One can make $\Dq_I(\Ri)$ a closed algebra by  
introducing the mapping
$\mu : \Dq_I(\Ri)\,\rightarrow\, 
       \hat{\Dq}(M)$. 
Here $\hat{\Dq}(M)$ is the algebra of 
$N\times N$ matrices and is spanned by  
the basis   
\be 
     \hat{\xi}^r = \mu(\xi^r) = 
    \left(
      \begin{array}{cc}
                 &   I_{N-r}  \\
             I_r &    
     \end{array}
   \right),  
   \quad r\in {\cal I}, 
\label{defop}     
\ee
where $I_n$ is the $n\times n$ unit matrix 
with the convention $I_{N+n}=I_n$.  
Notice that the algebra $\hat{\Dq}(M)$ 
is closed with the standard matrix multiplication, 
and $\hat{\cB}_I=\{\hat{\xi}^r \,\vert\, r\in \cI\}$ 
is a complete basis. 
Since the mapping $\mu$ is an isomorphism 
between $\Dq_I(\Ri)$ and $\hat{\Dq}(M)$,   
the multiplication in $\Dq_I(\Ri)$ is introduced as follows; 
upon denoting the multipliction as $\odot$, 
the product of two functions 
$\psi, \, \psi'\in \Dq_I(\Ri)$ is defined by  
\be
\psi(\xi) \odot \psi'(\xi) := \mu^{-1}\left( \mu(\psi)\mu(\psi')
    \right).    
\label{multi}     
\ee

Now the $q$-deformed algebra $\Dq_N(\RRn)$ 
of $q$-differentiable functions 
on $\RRn$ has been at hand, \ie.
\be
   \Dq_N(\RRn) \cong \cD(\cR) \otimes \Dq_I(\Ri;\odot),
\label{struc}   
\ee
where we have rewritten the internal sector 
as $\Dq_I(\Ri;\odot)$ 
in order to make the multiplication structure explicit.  
Thus, when $q$-differential structure is considered, 
the algebra of functions on $\Re_N$ 
is necessarily decomposed into two sectors: 
The first sector $\cD(\cR)$ is the standard 
(undeformed) algebra of differentiable functions,  
while the second sector $\Dq_I(\Ri;\odot)$, 
which is called the internal sector, is an algebra 
of $q$-differentiable functions 
with the multiplication \lq\lq$\odot$".  
At the same time, the base space $\Re_N$ has the tensor structure 
\be
  \Re_N = \cR \otimes \Ri \ni (x,\xi). 
\label{qR}   
\ee
Here, the space $\Ri$ is regarded as the internal space 
attached to every point of  the external space $\cR$.  
The geometrical structure of $\Re_N$ will be discussed 
in the next section.

\section{Quantum real lines}

We have obtained the algebras of $q$-differentiable functions:   
When the deformation parameter $q$ is {\it not} a root of unity,
we have $\Dq(\Re)$  on the base space $\Re$. 
In the case where  $q$ is at the $N$-th root of unity, 
the algebra is $\Dq_N(\Re_N)$ having the structure shown in 
(\ref{struc}) 
and the base space $\Re_N$ is 
expected to have the form in  (\ref{qR}).

Let us go ahead to the main task of our program, 
\ie, to investigate  geometrical structures of 
$\Re$ and $\Re_N$ 
and to propose quantum real lines. 
This final step is performed via the algebras 
$\Dq(\Re)$ and $\Dq_N(\Re_N)$.  
The quantum real line $\R_{\rm D}$ will be derived 
from $\Re$ in Section 4.1, 
and the lines $\qRN$ and $\vR$ will appear 
through  $\Re_N$ in Section 4.2.

\subsection{Quantum real line $\R_{\rm D}$ ;  
   $q$ is not a root of unity }

Notice first that, as we have seen in Section 3.1,  
the $q$-differential structure of the algebra $\Dq(\Re)$ 
appears as  the difference structure brought by $D_q$.  
It is, therefore, expected 
that the base space $\RRr$ 
has some discrete structure,  
although the variable $\xx$ has been 
treated as a continuous variable so far. 

One can find that the discrete structure appears in $\Re$ 
by choosing a point $x_{\vartheta_0}\in \Re$ arbitrarily. 
Upon the choice, 
the algebra $\Dq(\Re)$ reduces 
to the subalgebra $\Dq(\Re_0)$ 
where 
\be
  \Re_0\, :=\, \{ x_{\vartheta_n} \, \vert \, 
   \vartheta_n = \vartheta_0 + n\gamma, \,\, n\in \Z \,\}.
\label{pre}  
\ee
Recalling that $\xx = e^{i\vartheta}$, 
points in $\Re_0$ cover $S_1$ densely, 
since $\gamma$ is now an irrational number.   
Then, the algebra $\Dq(\Re)$ is just 
the direct sum of $\Dq(\Re_0)$ as 
\be
  \Dq(\Re) = \bigoplus_{0\leq 
      \vartheta_0 < \gamma} \Dq(\Re_0).
\label{preD}    
\ee
One immediately deduces that 
the base space $\Re$ is the union 
of all $\Re_0$ as 
\be
   \Re = \bigcup_{0\leq \vartheta_0<\gamma} \Re_0. 
\label{preR}   
\ee
It is important to notice that 
all the subalgebras are equivalent 
independently of the point $\vartheta_0$. 
Therefore, one has only to  pick up one subalgebra $\Dq(\Re_0)$ 
and $\Re_0$ as the base space,  
as far as the action of the operator algebra 
$\D(\Re)$ is considered.   
Now, the discrete space $\Re_0$ can be regarded 
as a candidate of the quantum real line.  
What we should  emphasize is that every point in $\Re_0$ 
is standard, \ie,  {\it local} point of null size. 
Actually,  the basis $\varpi$ of $\Dq(\Re)$ given in 
(\ref {basis}) guarantees that the Dirac's $\delta$-function 
exists in $\Dq(\Re)$.   
Therefore, the space $\Re$ and the subspace $\Re_0$ as well, 
consist of such standard points. 
This fact reads that 
a function $f\in \Dq(\Re_0)$ takes its values 
just on the  points $x_{\vartheta_n} \in \Re_0$ 
and, au each point, the function has sharp peak.

Let us define the $q$-deformed real line via $\Re_0$. 
To do this,  
we introduce the new variable $\zz$  through the relation  
\be
\zz_n = -i \,\log x_{\vartheta_n},  
\label{defzx}   
\ee 
and denote the space parameterized by 
$\zz_n$ as $\RD$, \ie,   
\be 
  \RD := \{ \zz_n \, \vert \,  n\in \Z \}. 
\label{RDD}     
\ee 
In order to obtain geometrical data of $\RD$,  
let us study $\Dg(\RD)$, 
the algebra of functions on $\RD$.
The difference operator $\Delta_\gamma$ 
with respect to the variable $\zz$  
is induced from $D_q$ as   
\be 
\Delta_\gamma= \frac{{\cal K}-1}{\gamma},  
   \label{difgamma}   
\ee
were we have assumed the relation 
$D_q=(D_q\zz) \Delta_\gamma$.  
The operator  ${\cal K}$ in (\ref{difgamma}) 
is the shift operator such as 
${\cal K}f(\zx)=f(\zx+\gamma)$ and,  therefore,   
the difference operator $\Delta_\gamma$ acts 
on a function $f \in \Dg(\RD)$ as  
\be
\Delta_\gamma  f(\zz) = 
     \frac{ f(\zz+\gamma) - f(\zz) }{ \gamma }.
\ee
Further,   
the key commutation relation between the coordinate $\zx$ 
and $\Delta_\gamma$ is calculated as  

\be
\chi \Delta_\gamma - \Delta_\gamma \chi = {\cal K}.   
\ee
Notice that the functional algebra  $\Dg(\RD)$ is closed 
under the action of  $\Delta_\gamma$,  \ie, 
all the functions in $\Dg(\RD)$ live just on $\R_{\rm D}$.  
We are at the stage to propose the base space $\RD$ as 
the quantum real line 
in the case where the deformation parameter $q$ is not 
a root of unity: \\[.3cm]
{ \rule{2mm}{2mm}  
   { \bf Proposition 1} : Quantum Real Line $\R_{\rm D}$ 
                                                    \\[.15cm] 
{\it  
The quantum real line $\R_{\rm D}$ 
for the case with generic deformation parameter $q$   
is a discrete space 
composed of an infinite number of points $\zz_n$ as}

\be
\RD = \{ \zx_n\, \vert \,  \zx_n =n\gamma +\zx_0,  
  \,\,  \,\,\,n \in \Z \,\}.
\ee 
 \\[.13cm]
It should be emphasized again that 
each point in $\RD$ is a localized object, 
\ie, $\zz_n$ are the standard  points in the Euclidean sense.  
The quantum real line $\RD$ is depicted in Fig.1.
\begin{displaymath}
\rule[.1cm]{10cm}{.38pt} \hskip-.365cm>\,\,\,\R_{\rm D}
  \hskip-9.8cm 
  \begin{array}{ccccccc}
     & & & & & \\[-.1cm]
    & \bullet 
    & \bullet 
    & \bullet 
    & \bullet 
    & \bullet 
    &                           \\
  \cdots & \zx_{n-2}  & \zx_{n-1} & \zx_n  
        & \zx_{n+1}  & \zx_{n+2} & \cdots
   \end{array} 
   \rule{3cm}{0pt}
\end{displaymath}

\vspace{.15cm}
\begin{center}
Fig.1  Quantum real line $\R_{\rm D}$
\end{center}
%
Thus, the quantum real line $\RD$ has the 
{\it minimal length} $\gamma$. 

Let us finally look at the limit 
$\gamma \rightarrow 0$, \ie, $q \rightarrow 1$.  
Since the interval between two adjacent points 
is $\gamma$,  
the quantum line $\RD$ becomes a continuous line. 
Further, $\R_{\rm D}$ becomes a smooth line,  
since the difference operator $\Delta_\gamma$
reduces to the standard  differential operator $\del_{\zz}$ 
with respect to the variable $\zx$.
These observations indicate that,  
in the limit $\gamma \rightarrow 0$, 
$\zx$ can be regarded as the standard real number 
and, therefore, 
$\R_{\rm D} \stackrel{\gamma \rightarrow 0}{\longrightarrow}\R $.

\subsection{Quantum real line  $\vR$ ; 
      $q$ is a root of unity}

Let us turn to the case 
where the deformation parameter $q$ is the $N$-th root of unity. 
We will investigate the base space $\Re_N$ 
via the algebra $\Dq_N(\Re_N)$ 
and propose the quantum real line for this case.  
Recalling that  the space $\Re_N$ is isomorphic t to 
the tensor product  of two sectors $\cR$ and $\Ri$ 
as shown in (\ref{qR}), 
we are going to investigate geometrical structures 
of these spaces $\cR$ and $\Ri$ separately 
in terms of the algebras $\cD(\cR)$ and $\Dq_I(\cS)$.

We have found in Section 3.2 that 
the first sector $\cD(\cR)$ is just the standard algebra 
of differentiable functions of $x\in \cR$. 
One then concludes that the base space $\cR$ is 
the standard real line, \ie, 
\be
   \cR=\R.
\label{concludion1}    
\ee
What to be made clear now is 
the structure of the second  sector $\Ri$.
Note first that $\cB_I=\{\xi^r\, \vert \, r \in \cI\}$ 
is a finite dimensional complete set.   
Namely, any function in $\Dq_I(\cS)$ can be 
expanded uniquely by $\xi^r, \, r\in \cI$   
and, therefore, the following function,  referred to it as 
\lq\lq wave function of a point", 
can be defined in $\Dq_I(\cS)$:     \\[.3cm]
 \rule{2mm}{2mm}   {\bf Definition}: \quad 
Wave function of a point in $\Ri$

For a point $\xi'= e^{i\vartheta'} \in \Ri$, 
the folloowing function on  
the space $\Ri$ $(-\pi \leq \vartheta < \pi)$ 
can be defined as the wave function of the point 
$\xi'$, 
\be
    \Psi_{\vartheta'}(\vartheta) := \frac{1}{\sqrt{2\pi N}}  
    \sum_{r\in {\cal I}} \left( \xi^{\ast} \right)^r \cdot
      \left( \xi' \right)^r 
      = \frac{1}{\sqrt{2\pi N}}
  \sum_{ r\in {\cal I}} e^{ir(\vartheta'-\vartheta)}. 
\label{delta}     
\ee
\\[.13cm] 
Notice that $\Psi_{\vartheta'}(\vartheta)$ is normalized as 
\be
\int_{-\pi}^\pi 
    \left\vert
    \Psi_{\vartheta'}(\vartheta) 
        \right\vert^2 \, {\rm d} \vartheta=1.     
\ee   
Then, one can regard 
\be
    \rho_{\vartheta'}(\vartheta) := \left\vert 
      \Psi_{\vartheta'}(\vartheta) 
    \right\vert^2
\label{density}  
\ee
as the density of the point $\xi'$. 
With a finite $N$, 
the density (\ref{density}) indicates that 
the point $\xi'$ spreads, \ie,  
fluctuates around the position $\vartheta=\vartheta'$ 
by the quantum effect. 
In this sense, such a  point is called 
a \lq\lq fuzzy point.   
We should remark the feature in   the limit $N\rightarrow \infty$. When $N$ goes to infinity, the wave function 
$\Psi_{\vartheta'}(\vartheta)$ has a 
$\delta$-functional peak only at $\vartheta = \vartheta'$, 
and  so, the point $\\xi'$ locates at 
the position $\vartheta = \vartheta '$ 
with infinite density.

We further remark that   
the algebra $\Dq_I(\Ri)$ is just the direct sum of 
an infinite number of subalgebras 
which are closed under the action of the difference operator 
$\nabla_q$. 
As we have done in the preceding subsection, 
one finds  
\be
    \Dq_I(\Ri) = \bigoplus_{0\leq \vartheta_0<\frac{2\pi}{N}}
    \cA_N(\Rin_{\vartheta_0}).
\label{IntSpa}   
\ee
Let us observe  a subalgebra $\cA_N(\Rin_{\vartheta_0})$ 
in detail and, then,  
make the structure of $\Rin_{\vartheta_0}$ clear. 
Taking it into account that 
the difference operator $\nabla_q$ 
generates the finite displacement   
$\xi\,\rightarrow\,q\xi$ 
with $q=e^{i\frac{2\pi}{N}}$, 
one sees that $\cA_N(\Rin_{\vartheta_0})$ is 
the space of functions on $N$ fuzzy points as  
\be
  \cA_N(\Rin_{\vartheta_0}) = \{ \psi(\xi_n) \, \vert \, 
         \xi_n = e^{ i\frac{2\pi n}{N}}\xi_0 , 
         \quad n=0,1,\cdots, N-1
             \}.   
\ee 
Namely, the internal space $\Rin_{\vartheta_0}$  
is composed of $N$ fuzzy points,   
\be   
   \Rin_{\vartheta_0}= \{ \vartheta_n\, \vert \, 
     \vartheta_n= \vartheta_0 + 
        \mbox{$\frac{2\pi}{N}\,n$}, \,\, n=0, 1, \cdots, N-1 
                           \,   \},  
\ee
where, instead of $\xi_n$,   
the variables $\vartheta_n$ 
defined by $\xi_n= e^{i\vartheta_n}$ is used 
for the parametrization of $ \Rin_{\vartheta_0}$. 
As we have explained, a point $\vartheta_n$ 
extends over the whole $\Rin_{\vartheta_0}$ 
according to $\Psi_{\vartheta_n}(\vartheta)$,   
and the distribution of the $N$ 
fuzzy points along $\Rin_{\vartheta_0}$ is given by 
the superposition   
\be 
   \sum_{n=0}^{N-1} \, \Psi_{\vartheta_n}(\vartheta). 
\ee  

Now, putting (\ref{qR},\ref{concludion1}) together, 
we are at the stage 
to investigate the total space $\qRN$ 
which is a subspace of  $\Re_N$,  \ie, 
\be
   \qRN := \R \times \Rin_{\vartheta_0} 
   \ni (x,\, \vartheta_n). 
   \label{titspa}   
\ee 
The total algebra $\cD(\R) \otimes \cA_N(\Rin_{\vartheta_0})$ 
is built on $\qRN$. 
Before proposing $\qRN$ as a quantum real line, 
we should make clear the relationship 
between the external space $\R$ and 
the internal space $\Rin_{\vartheta_0}$. 
The difference operator $\nabla_q$ 
maps a point in $\qRN$ as  
\be
   \nabla_q\, : \, (x,\vartheta_n) \quad 
\llmapsto \quad
              (x, \vartheta_{h+1})
\label{mapnabla}    
\ee
where $(x,{\cal P}) \mmapsto (x,{\cal Q})$ 
stands for the mapping 
from the point $(x,{\cal P})$ to the points 
$(x,{\cal P}) \cdots (x,{\rm Q})$ 
along $\Rin_{\vartheta_0}$.  
By taking the $N$-th power of the mapping 
(\ref{mapnabla}), one finds
\be
  \nabla_q^N\,: \, (x, \vartheta_n) 
    \quad \llmapsto \quad  (x, \vartheta_{n+N}). 
\ee   
As far as only the internal space $\Rin_{\vartheta_0}$ 
is considered, 
$\nabla_q^N$ brings a point $\vartheta_n$ 
to the same position, 
since $\vartheta_{n+N}=\vartheta_n +2\pi$.
However, once we look over the total space 
(\ref{titspa}), 
the equivalence $D_q^N/[N]! \equiv \del_x$ 
indicates 
\be
(x,\vartheta_n+2\pi) =(x+\epsilon, \vartheta_n)  
\ee   
with an infinitesimal number $\epsilon \in \R$.  
Thus, the external space and the internal space 
are not independent of each other but related as follows:   
Starting from a point $x\in \R$ 
and moving along $\Rin$, 
one reaches the point $x+\epsilon \in \R$.

Now, we have understood the structure of $\qRN$, 
and it is the time to propose $\qRN$ as a 
quantum real line when $q$ is a root of unity. \\[.3cm] 
\rule{2mm}{2mm}  {\bf Proposition 2} : Quantum Real Line $\qRN$
                                                  \\[.1cm] 
{\it The quantum real line $\qRN$ 
for the case with the deformation parameter $q$   
at the $N$-th root of unity  
is given by the product of
 $\R$ and $\Rin_{\vartheta_0}$ as} 
\bee
    \qRN &=& \R \times \Rin_{\vartheta_0}  
                     \nonumber
\\[.15cm]
     &=& \{ (x,\vartheta_n) \, \vert \, 
    (x,\vartheta_{n+N}) =(x+\epsilon, \vartheta_n)
    \}   \label{QRL2}   
\eee     
{\it where $\epsilon$ is an infinitesimal real number
and $n =0,1, \cdots, N-1$}. 

\vspace{.2cm}

From  the viewpoints of the quantum real line $\qRN$, 
each real number $x$ is furnished with   the internal space 
$\Rin_{\vartheta_0}(x)$ which is composed of $N$ fuzzy points 
$\vartheta_n,\, n=0, \cdots, N-1$.  
On the other hand, the next point of $\vartheta_{N-1} $, 
\ie,  $\vartheta_{N}$ belongs to 
$\Rin_{\vartheta_0}(x+\epsilon)$ on 
$x+\epsilon$ with an infinitesimal $\epsilon\in \R$.  
Thus, the internal space $\Rin_{\vartheta_0}(x)$ 
connects two infinitesimally separated 
numbers $x$ and $x+\epsilon$.  
In this sense, the fuzzy 
internal space $\Rin_{\vartheta_0}$ will be called,   
hereafter, an {\it infinitesimal structure}.

It is interesting  to  observe the infinitesimal structure 
from another viewpoint 
and propose another description 
of the quantum real line. 
To this end, we introduce the map 
$\check{\pi} : \{x\}\times \Rin_{\vartheta_0}(x) \rightarrow
\cM(x)$ such as 
\be
   \check{\pi}\,:  \,\, (x, \vartheta_n)\,\, \longmapsto 
   \,\, \cx_n:= x \cp  n \cep, \quad n =0,1,\cdots, N-1, 
\label{SN}    
\ee
where $\cep$ is some constant and 
the symbol $\cp$ stands for the sum operation 
in the space $\cM(x)$. 
Then, the space $\cM(x)$ which is defined for each $x\in\R$ 
is written  explicitly as 
\be
\cM(x) = \{ \cx_n\, \vert \, n=-0,1,\cdots,N-1, \,\,\, 
     \cx_0\equiv x \}. 
\label{defM}    
\ee
It should be emphasized that every point $\cx_n$ is fuzzy, \ie, 
the space $\cM(x)$ is a fuzzy space as well as 
$\Rin_{\vartheta_0}$.

We have now prepared for giving 
the proposition of 
another quantum real line 
when $q$ is the $N$-th root of unity:  \\[.3cm]
\hfill\rule{2mm}{2mm}  {\bf Proposition 3} : Quantum Real Line $\vR$ 
                                                  \\[.1cm] 
{\it The qantum real line $\vR$ 
is given by the union of $\cM(x)$ as},  
\be
\vR  = \bigcup_{x\in\R} \,\, \cM(x). 
\label{Rspace}      
\ee
{\it where $\cM(x)$ is the fuzzy space 
embedded between two real numbers $x$ and $x+\epsilon$ 
with an infinitesimal real number $\epsilon \in \R$}.
 \\[.13cm] 
The quantum real line $\vR$ is shown in Fig.2. 
\bee
&&              \nonumber \\
&&\cdots  \,\,
\overbrace{ \rule{0pt}{.5cm} 
    \bullet \hskip.7cm  \circ \hskip.7cm  \circ \quad \cdots \quad 
        \circ 
             \,  }^{\cM(x)} \quad \,\,\,
\overbrace{ \rule{0pt}{.5cm} \,
  \bullet \hskip.7cm \circ \hskip.7cm  \circ \quad \cdots \quad \circ 
             \,  }^{\cM(x+\epsilon)} \,\, \cdots 
   \nonumber \\
&& \rule{.8cm}{0pt} x \hskip.7cm \, \cx_1 
    \rule{2.15cm}{0pt} \cx_{N-1} \hskip.35cm  
    x\!+\!\epsilon 
       \nonumber   \\[-.15cm]
&& \rule{.85cm}{0pt} \longleftarrow \,\quad 
      \mbox{{\small infinitesima}l}
           \,\quad \longrightarrow \nonumber 
\eee
\begin{center}
Fig.2 \quad  Quantum real line  $\vR$
\label{fig1}      
\end{center} 
where the bullets $\bullet$ stand for the 
\lq\lq standard" real numbers $\R$, 
while the circles $\circ$ represent numbers in 
$\vR\setminus \R$. 
It is natural to call the space $\cM(x)$ as an  
\lq\lq infinitesimal structure",   
since $\cM(x)$ lies between $x$ and $x+\epsilon$ 
as shown in Fig. 2.    
It is quite important to notice the status of 
the number $\epsilon$. 
As long as we observe $\epsilon$ within the framework of $\R$,  
it is infinitesimally small. 
However, from the viewpoint of $\vR$,  
it is no longer an infinitesimal. 
Note also that the constant $\ce$ 
appeared in  (\ref{SN}) 
is not a real number but an element of $\vR\setminus\R$.


We have obtained two quantum real lines $\qRN$ and $\vR$ 
when $q$ is the $N$-th root of unity. 
Some remarks are in order.
They are, of course, identical with each other 
under the isomorphism $\check{\pi}$.  
The quantum line $\qRN$ has been introduced  as  
a two-dimensional space. 
One of the dimensions corresponds to 
the standard real line $\R$, 
\ie, noncompact and continuous space.   
The other is $\Rin_{\vartheta_0}$ which 
is compact and fuzzy $N$-point space. 
We have called the extra dimension $\Rin_{\vartheta_0}$ 
the infinitesimal structure.
On the other hand, 
we have defined $\vR$ as {\it an extension }of the standard 
real numbers $\R$ by introducing 
the numbers with \lq\lq $\check{\rule{5mm}{0pt}}$".  
Due to the extension,    
the quantum line $\vR$ can be viewed 
as a one-dimensional space.   
In  $\vR$,  
the infinitesimal structure $\cM(x)$ 
appears between two adjacent real numbers 
$x$ and $x+\epsilon$. 
We should stress that, 
only for the case when 
these numbers $x$ and $x + \epsilon$ are understood 
just in the framework of $\R$,   
the interval between them 
is regarded as of infinitesimally small.

\section{Concluding Remarks}

In this paper, we have investigated  
 possible deformations  
of the real numbers by taking quantum effects into account 
and proposed quantum real lines.  
Upon noticing that the Heisenberg algebra $\cL$ 
spanned by $a,\,\ad$ is represented by 
 the algebra of differentiable functions on $\R$,
our discussions have started from the deformation of 
$\cL$ 
according to the Moyal deformation procedure. 
We have introduced an internal space $\cT$ and  
a functional algebra $\cA(\cT;\ag)$ on $\cT$.  
The multiplication in $\cA(\cT, \ag)$ is 
the Moyal product $\ag $ with respect to 
the deformation parameter 
$\gamma$ and, then,  
$\cA(\cT; \ag)$ is the noncommutative torus.  
Although the algebra $\cL\times \cA(\cT; \ag)$   can be regarded as  
a $q$-deformed Heisenberg algebra with the deformation parameter 
$q = e^{I\gamma}$,  
we have further derived  the  algebra 
$\cLLq:=\cL\times 
\cAA(\cTT)$ 
from $\cL \times \cA(\cT; \ag)$ by   
reducing the base space $\cT$ to the one-dimensional space $\cTT$. 
The crucial point here is that, 
due to the reduction $\cT \rightarrow \cTT$, 
the functional algebra $\cA(\cT;\ag)$ becomes 
the operator algebra $\cAA(\cTT)$.   
We have treated  $\cLLq$  as our $q$-deformed 
Heisenberg algebra.  

The algebra $\cLLq$ has been represented by  
the algebras of $q$-differentiable functions 
on the  base spaces  $\RRr$ and $\RRn$,   
\ie,  the algebra $\Dq(\RRr)$ for the case 
where $q$ is not a root of unity 
and $\Dq_N(\RRn)$ for the case 
where $q$ is the $N$-th root of unity.  
Focussing on the $q$-differential structures of these algebras,  
the geometrical structures of the base spaces 
$\RRr$ and $\RRn$ have been deduced.  
Finally, we have proposed the quantum real line $\R_{\rm D}$   
from $\RRr$ and lines $\qRN$ and $\vR$ from $\RRn$.

Let us summarize the features of these quantum lines.  
\begin{itemize}  
\item The quantum real line $\R_{\rm D}$\\
  It is a discrete space composed of 
 $\cX_n, \, n\in \Z$.  
A quantum real number in $\R_{\rm D}$ is localized upon a spot 
  and distance between two adjacent spots 
  is specified  by $\gamma$.  
\item The quantum real line $\qRN$ \\
      It is a two-dimensional object as 
      $\qRN=\R \times \Rin_{\vartheta_0} 
        \ni (x, \vartheta_n)$ 
       where $\Rin_{\vartheta_0}$ is composed of 
       $N$ fuzzy points.   
      The fluctuation of a point $\vartheta_n $ 
            along the internal space $\Rin_{\vartheta_0}$ 
            is described by the wave function 
      $\Psi_{\vartheta_0}(\vartheta)$ given in 
     (\ref{delta}). 
    The important  fact is that 
     moving around $\Rin_ {\vartheta_0}$ 
      generates an infinitesimal displacement along $\R$.  
\item The quantum real line $\vR$ \\
     It is isomorphic to $\qRN$. 
     However, in defining $\vR$, 
     we have extended  the concept of the standard 
     real number, \ie,  
       $\vR$  is composed of the extended real numbers  
      $\cx_n= \cx_0 \cp n\ce, \,  n=0,1,\dots, N-1$  where 
     $\cx_0 \equiv x \in \R $ and $\ce \in\!\!\!\!\!/ \,\, \R$ 
    is the unit of $\vR$.  
    Thus, 
    each real number $x \in \R$ possesses 
   the infinitesimal structure 
   $\cM(x)=\{ \cx_0, \cx_1, \cdots, \cx_{N-1} \} $ 
   between $x$ and $x+\epsilon$. 
     An important fact is that $\epsilon$ is treated as 
     an infinitesimal within $\R$, 
     while it is not within $\vR$.  
    The $N$-point space $\cM(x)$ 
   is a fuzzy space as well as $\Rin_{\vartheta_0}$.  
    the infinitesimal structure of $x$.  
\end{itemize} 

It is interesting to consider the {\it classical} 
limit $q \rightarrow1$ in each quantum line. 
For the line $\R_{\rm D}$, 
the limit corresponds to the situation 
where the parameter $\gamma$ is so small that 
the distance $\cX_{n+1}-\cX_n \sim \gamma$ 
shrinks to zero. 
Furthermore, in the limit, the difference operator 
$\Delta_\gamma$ returns to  
the standard differential operator. 
Therefore, $\R_{\rm D}$ goes back to a continuous 
and smooth one-dimensional line, \ie,  
the standard real line $\R$. 
In other words,  the limit corresponds to  the situation 
where the wave length $\lambda$ to observe $\R_{\rm D}$ is 
so long $\lambda \gg \gamma$ 
that the discreteness of the quantum line  is not visible.

The classical limits of the lines 
$\qRN$ and $\vR$ are to be noticed.  
In these lines, 
$q \rightarrow1$ is realized 
by taking $N=1$ or $ N\rightarrow\infty$.    
When $N=1$, 
$\cM(x)$ becomes the one-point space 
$\cM(x)\vert_{N=1} =\{\cx_0=x\in\R \}$,  
\ie, the infinitesimal structure vanishes and  $\vR=\R$.  
On the other hand, when $N \rightarrow \infty$,  
the space $\cM(x)$ becomes a continuous space. 
Similarly, for the line $\qRN$, 
the internal space $\Rin_{\vartheta_0}$  
becomes 
 a one-point space when $N=1$ and the smooth circle $S_1$ 
wheen $N \rightarrow \infty$.  
However, for the limit $N\rightarrow \infty $, 
there remains a room to discuss 
whether $\cM(x)$ and $\Rin_{\vartheta_0}$  
expand infinitely large 
or keep small.  
If $\cM(x)$ extends to infinity, 
then $\cM(x)$ itself is raised to  $\R$.  
On the contrary, 
if the size of $\Rin_{\vartheta_0}$ remains small, 
$\vR \rightarrow \R \times S_1$, \ie, 
two-dimensional cylinder appears!

Let us end this paper by giving a comment 
on another limit for $\vR$.  
It should be noticed that 
it is possible to take 
the limit 
$\ce \rightarrow0$ with keeping $N$ finite. 
In this case, $\epsilon$ also goes to 0. 
Then, it seems that $\vR$ reduces to $\R$. 
However, there is a crucial difference, \ie,
upon defining  
\be
  \widehat{\R}_N  := \lim_{\ce\rightarrow 0} \vR, 
\ee   
$\widehat{\R}_N$ is also an extension of $\R$.
Each point in $\widehat{\R}_N$ is degenerated by 
$N$ number of points 
$\cx_0, \cx_1, \cdots, \cx_{N-1}$. 
In other words, 
$\widehat{\R}_N$ is regarded as a bundle of 
$N$ real lines.  
What  happens to a function $\Phi$ on $\vR$ 
in this llimit?  
Since $\vR$ looks like a bundle of $N$ real lines 
and $\cx_r$  becomes a point on the $r$-th line,  
one finds the reduction 
$ \Phi(\cx_r)  \rightarrow \phi_r(x)$. 
Here,  $\phi_r(x)$ is regarded as a function 
living on the $r$-th real line.  
Thus, in the limit $\epsilon \rightarrow 0$, 
function $\Phi$ can be viewed 
as a unification of $N$ functions $\phi_i$ 
on $\R$ as   
$\Phi \rightarrow 
  \phi_0 \oplus \phi_1 \oplus \cdots \oplus \phi_{N-1}$. 

An important application will be found when  $N=2$.  
In this case, the space $\cM(x)$ is the two-point 
space such as $\cM(x)=\{ \bullet,\,\circ\}$.  
In the limit $\epsilon \rightarrow 0$,  
$\widehat{\R}_2$, \ie, 	
a pair of two real lies apers. 
It is interesting to expect  that 
supersymmetric models are described in our formalism. 
In explicit, one of the real lines in $\widehat{\R}_2$ 
corresponds to the space where bosons $\phi_B$ live 
and on the other line,  
fermions $\psi_F$ live. 
Namely, field $\Phi$ can be regarded as a superfield  
$\Phi \rightarrow \phi_B(x) \oplus \psi_F(x)$. 
The above prospect follows from the facts:  
Actions of the generator of supertransformation  
$\delta_S$ move, e.g., $\phi_B(x)$ as 
$ \phi_B(x) \stackrel{\delta_S}{\longrightarrow} \psi_E(x) 
   \stackrel{\delta_S}{\longrightarrow} \phi_B(x+\epsilon)$.  
Similarly, in our model with $N=2$, 
we have seen the movements 
$(x, \circ) \stackrel{D_q}{\longrightarrow} (x, \bullet) 
    \stackrel{D_q}{\longrightarrow}
     (x+\epsilon, \circ)$. 
Furthermore,  the author has shown the equivalence 
between the quantum universal enveloping algebra 
$\qslc$ with $q$ at the 2-nd root of unity and 
the supersymmetric algebra ${\rm Osp}(2\vert1)$ \cite{TS2}. 
Thus, one expect that quantization of geometry is the origin of 
the supersymmetry. 
 The investigation will appear elsewhere.

\vspace{1cm}


\noindent
{\bf Acknowledgments} \\[.15cm]
The author would like to thank Dr. T. Araki and T. Kogisso 
for valuable discussions and comments.


\end{document}